\documentclass[aps,prb,twocolumn,groupedaddress,floatfix,showpacs]{revtex4}   %%%%%%%%%%%%%%%%%%%%%%%%%%%%%%%%%%%%%%%%%%%%%%%%%%%%%%%%%%%%%%%%%%%%
\usepackage[dvipdf]{graphicx}

\begin{document}
%  \pagestyle{plain}
%\preprint{}
\title{Role of local fields in the optical properties of silicon nanocrystals using the tight binding approach}
\author{F. \surname{Trani}}
\email{trani@na.infn.it}
\affiliation{Dipartimento di
Scienze Fisiche, Universit\`a di Napoli ``Federico II,''\\ Complesso Universitario Monte S. Angelo, Via Cintia, I-80126 Napoli, Italy
}
\author{D. \surname{Ninno}}
\author{G. \surname{Iadonisi}}
\affiliation{Dipartimento di
Scienze Fisiche and Coherentia CNR-INFM, Universit\`a di Napoli ``Federico II,''\\Complesso Universitario Monte S. Angelo, Via Cintia, I-80126 Napoli, Italy
}

\begin{abstract}

The role of local fields in the optical response of silicon nanocrystals is analyzed using a tight binding approach. 
Our calculations show that, at variance with bulk silicon, local field effects dramatically modify the silicon nanocrystal optical response. An explanation is given in terms of surface electronic polarization and confirmed by the fair agreement between the tight binding results and that of a classical dielectric model. From such a comparison, it emerges that the classical model works not only for large but also for very small nanocrystals. Moreover, the dependence on size of the optical response is discussed, in particular treating the limit of large size nanocrystals.

\end{abstract}

\pacs{73.63.Kv,78.67.Bf,73.43.Cd}

\maketitle

Silicon nanocrystals (Si-\textit{nc}) constitute very promising building blocks for the next generation of optoelectronic devices. Since the discovery of a strong optical activity of Si-\textit{nc}, \cite{wilson93} their fabrication techniques have been much improved. Nowadays, small ($< 2$ nm) Si-\textit{nc} samples with narrow size distributions and high quantum yields are produced by chemical synthesis.\cite{wilcoxon99,warner05,jurbergs06} An open discussion in literature is on the influence of the passivant on their optical response. Recent experiments have shown that new kinds of passivation by organic molecules may open new scenarios for enhancing the Si-\textit{nc} quantum yield.\cite{warner05,jurbergs06} The passivation and the background in which Si-\textit{nc} are synthesized cannot therefore simply be ignored, and their role still must be carefully examined. From a theoretical point of view, some work has been done in analyzing the effects of passivant species in silicon nanocrystals. \cite{puzder02} But very few papers discuss the contribution of the background from an atomistic point of view.
Indeed, standard calculations are based on the independent particle approach, and they fully neglect the presence of a background embedding medium, treating the nanocrystal as an isolated object. In this paper, using a tight binding framework, we show that local field effects (LFEs)\cite{sigle} deeply influence the Si-\textit{nc} optical response. We compare our results (i) with a dielectric model based on classical electrostatics, (ii) with first principles calculations, and (iii) with experimental data. In the following, the dielectric model and the atomistic approach are illustrated. The results are then discussed.

There are two remarkable differences between the optical response of bulk Si and Si-\textit{nc}. The first is the quantum confinement effect (QCE). Quantum confinement has given for years a qualitative and often quantitative description of the Si-\textit{nc} photoluminescence blueshift upon decreasing the nanocrystal size, and it has been widely studied using both semiempirical and first principles computational tools.\cite{trani05} The second difference, whose importance, in our opinion, has always been underestimated, is the surface polarization effect (SPE). When a dielectric structure is embedded into a background characterized by a different dielectric constant, its optical response can dramatically change. Indeed, as it is well known from the classical electrostatics, the charge accumulation due to the dielectric mismatch across the surface causes the macroscopic field inside to be very different from the external field.\cite{landau}
Numerous definition of the dielectric constant of a nanocrystal can be given.\cite{wang94,lannoo95,trani05} 
We follow a procedure, in which, starting from the polarizability $\alpha$ of a Si-\textit{nc} embedded into a background with dielectric constant $\varepsilon_0$, an \textit{effective} dielectric constant is defined as $\varepsilon_{\text{eff}}=\varepsilon_0 +4\pi \alpha/\Omega$ ($\Omega$ is the structure volume). $\varepsilon_{\text{eff}}$ is a dielectric constant \textit{with SPEs included}. In the case of noninteracting spheres, embedded into a background with dielectric constant $\varepsilon_0$, the dielectric constant is given by\cite{landau,nota1}
\begin{equation}
   \label{eq:classical}
   \varepsilon_{\text{eff}} =  \varepsilon_0 \left[\frac{4\varepsilon _{\text{Si}} - \varepsilon_0}{\varepsilon _{\text{Si}}+ 2\varepsilon_0}  \right],
\end{equation}
where $\varepsilon_{\text{Si}}$ is the bulk Si dielectric constant.
We refer to this way of calculating SPEs (that leads to Eq. (\ref{eq:classical}) for spheres) as to the classical model (CM), in that QCEs are neglected, and classical electrostatics is used. The total dielectric constant $\varepsilon_m$ of the Si-\textit{nc} plus the background mixture is obtained by a weighted average between Eq. (\ref{eq:classical}) and the background dielectric constant. Introducing the Si-\textit{nc} filling ratio $f$, the average is simply given by
\begin{equation}
 \label{eq:averaging}
  \varepsilon_{m} = f \varepsilon_{\text{eff}} + (1-f)\varepsilon_0.
\end{equation}
With the explicit substitution of $\varepsilon_{\text{eff}}$, Eq. (\ref{eq:averaging}) gives the linearized Maxwell-Garnett effective medium equation. \cite{deleruebook} An enhancement of the model is obtained when a size dependent dielectric constant is used as $\varepsilon_{\text{Si}}$. We call this latter the semiclassical model (SCM), in that, even if SPEs are calculated classically, the model takes into account QCEs. We stress here that $\varepsilon_{\text{Si}}$ \textit{does not include SPEs}, and cannot be directly compared to experiments, unless SPEs are negligible.

The available quantity in experiments is $\varepsilon_m$, the mixture dielectric constant.
Therefore the comparison with the experimental data can be done following three steps.
(1) The starting point is a dielectric constant, which either includes (SCM) or not (CM) the QCEs. (2) SPEs are introduced in a classical way, using Eq. (\ref{eq:classical}) in the case of spherical non interacting Si-\textit{nc}. (3) A weighted average with the background dielectric constant is done, introducing the Si-\textit{nc} filling factor. Points (2) and (3) constitute the basis of effective medium models, that in the case of noninteracting particles (low filling factors), reduce to the Maxwell-Garnett equation.
Equation (\ref{eq:classical}) can be generalized to a frequency dependent dielectric constant. In this case $\varepsilon_{\text{Si}}$ is a complex quantity, and from $\varepsilon_{\text{eff}}$ the absorption cross section can be derived. The whole procedure goes under the name of dielectric model, since the materials are treated as continuous dielectric media. The model is commonly used to compare theoretical results and experimental data.\cite{kovalev95} The model is expected to fail either for small or for nanocrystals with a complex shape. For this reason it is important to calculate SPEs within an atomistic quantum mechanical framework, that reduces to the dielectric model for large size, regular shaped Si-\textit{nc}.
Moreover, an atomistic scheme is absolutely needed for the study of the effects on the optical response due to both the presence of inhomogeneities on the surface, and the chemical nature of the passivants.\cite{puzder02,gatti05} In the following we describe how SPEs are calculated using quantum mechanical atomistic theories.
% 
% \begin{figure}
% 	\centering
% 	\includegraphics[width=.4\textwidth]{fig1.eps}
% 	\caption{(Color online) Static dielectric constant calculated for a set of increasing size Si-\textit{nc} using different levels of approximation: RPA, blue-crosses; RPA+LF, black-squares; semiclassical results, red-plus symbols; classical model, green solid line. The lines are guides for the eyes.}
% 	\label{fig:fig1}
% \end{figure}

For years, dielectric constant calculations have been performed within an independent particle scheme, using the so called random phase approximation (RPA). \cite{wang94,trani05}
Independent particle RPA takes into account the QCEs, but it neglects the electron-hole interaction\cite{leung97} and especially LFEs. These are related to the inhomogeneous nature of the electron density induced by an external field. While in crystalline Si LFEs are basically due to the polarization of Si-Si bonds giving a negligible contribution to the dielectric constant, in Si-\textit{nc} they are very important due to the presence of an interface between each nanocrystal and the background.\cite{trani06} In such a case, LFEs are mostly due to SPEs, as it has been recently shown in the case of silicon quantum wires.\cite{bruneval05} Due to the huge computational task, only recently LFEs have been calculated using first principles schemes.\cite{bruneval05,gatti05,sottile05}

We use a semiempirical tight binding scheme which gives the unique
opportunity of studying structures made of hundreds of Si atoms. Details on the method and the approximation used for the position matrix elements can be found elsewhere.\cite{trani05,delerue97} It is just worth mentioning that we make use of an $sp^3$ third nearest-neighbor parametrization, which provides for bulk silicon a good description of both the band structure and of the optical response.\cite{trani05}
Using the diagonal approximation for the position matrix elements (see Ref. \onlinecite{trani05} and references therein), we obtain the following expression for the frequency dependent dielectric tensor:\cite{hanke74}
\begin{equation}
\label{eq:tb}
\varepsilon _{\beta,\gamma} \left(\omega\right) = 1 - \frac{4\pi e^2}{\Omega} \sum _{i,j }\mathbf{R}_i^{\beta} S\left(\omega\right)_{ij} \mathbf{R}_j^{\gamma}.
\end{equation}
where $\Omega$ is the Si-\textit{nc} volume, $\mathbf{R}_i^{\beta}$'s are the positions of atoms composing the structure, $\beta$ and $\gamma$ are Cartesian components, $S_{ij}$ is the real space screened polarization matrix. $S$ is defined by $S=P\varepsilon^{-1}$ (matrix multiplication), where $P_{ij}$ and $\varepsilon_{ij}$ are the polarization and the dielectric matrix in a tight binding representation, that we calculate following Ref. \onlinecite{delerue97}. When $S$ is substituted by $P$ (that is, when the polarization is unscreened), Eq. (\ref{eq:tb}) gives the standard independent particle RPA expression.

We thus study the Si-\textit{nc} optical response using one of the following approximations:
(i) RPA (QCEs are taken into account, SPEs neglected),
(ii) RPA+LF (QCEs and LFEs are calculated within the tight binding framework),
(iii) SCM (QCEs are calculated using the RPA approximation, SPEs using the dielectric model),
(iv) CM (QCEs are neglected, $\varepsilon_{\text{Si}}$ is the bulk Si dielectric function, SPEs are calculated using the dielectric model).

\begin{figure}[t]
 	\includegraphics[width=0.4\textwidth]{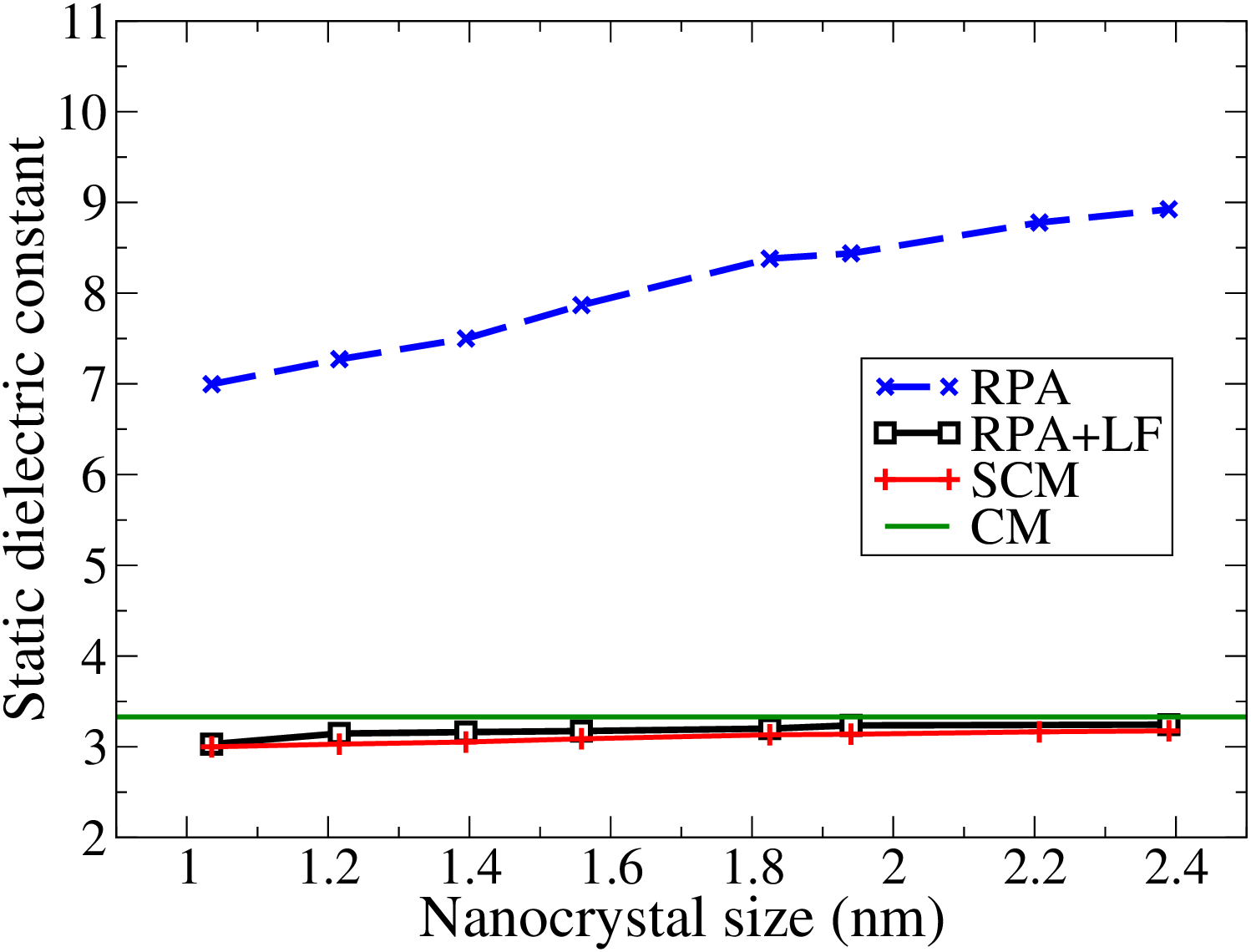}
	\caption{(Color online) Static dielectric constant calculated for a set of increasing size Si-\textit{nc} using different levels of approximation: RPA, blue crosses; RPA+LF, black squares; semiclassical model, red plus symbols; classical model, green solid line. The lines are guides for the eyes.}
	\label{fig:fig1}
\end{figure}

In Fig. \ref{fig:fig1} we show the static dielectric constant for spherical hydrogenated Si-\textit{nc}, as a function of the size. RPA, RPA+LF, SCM, and CM results are compared. First we note that RPA+LF calculations (black squares) well agree with the SCM approach (red plus symbols), in which SPEs are included into the RPA dielectric constant using Eq. (\ref{eq:classical}). It is remarkable that the semiclassical model gives an extremely accurate result not only in the large size region, where it is expected to work well, but also for very small Si-\textit{nc}. Second, when LFEs are taken into account, the dependence on size of the effective dielectric constant becomes very weak. Figure \ref{fig:fig1} shows that for the RPA+LF calculation there is a relative change of about 3\%  when the size changes from 1.2 to 2.4 nm, against 20\% of the RPA curve. Moreover, the convergence to the large size limit is much faster, since it is reached already for small Si-\textit{nc}. This means that in experimental situations in which SPEs are important, SPEs play a role which is the opposite of QCEs, in that their contribution is size independent. For large size structures (dimensions greater than the Si exciton Bohr radius), QCEs are negligible and both RPA+LF and SCM tend to the CM limit. In Fig. \ref{fig:fig1}, for the CM calculation, the experimental value $\varepsilon_{\text{Si}}=11.4$ is used in Eq. (\ref{eq:classical}). A remarkable information emerging from Fig. \ref{fig:fig1} is the different behavior of RPA and RPA+LF results upon increasing the nanocrystal size. While RPA tends to the bulk Si dielectric constant, RPA+LF does not. This means that LFEs (that is SPEs) lead to a significant suppression of the static dielectric constant \textit{for each nanocrystal size}. This is due to the fact that, at variance with the microscopic QCEs, SPEs give a \textit{macroscopic} contribution. For this reason we could say that the effective dielectric constant is a physically different property than the usual (RPA) Si-\textit{nc} dielectric constant, and it is necessary to distinguish between calculations in which SPEs are either neglected or included. It is remarkable that SPEs comes out \textit{directly} from the standard linear response theory, and there is no need of any further step.

\begin{figure}[t]
 	\includegraphics[width=.4\textwidth]{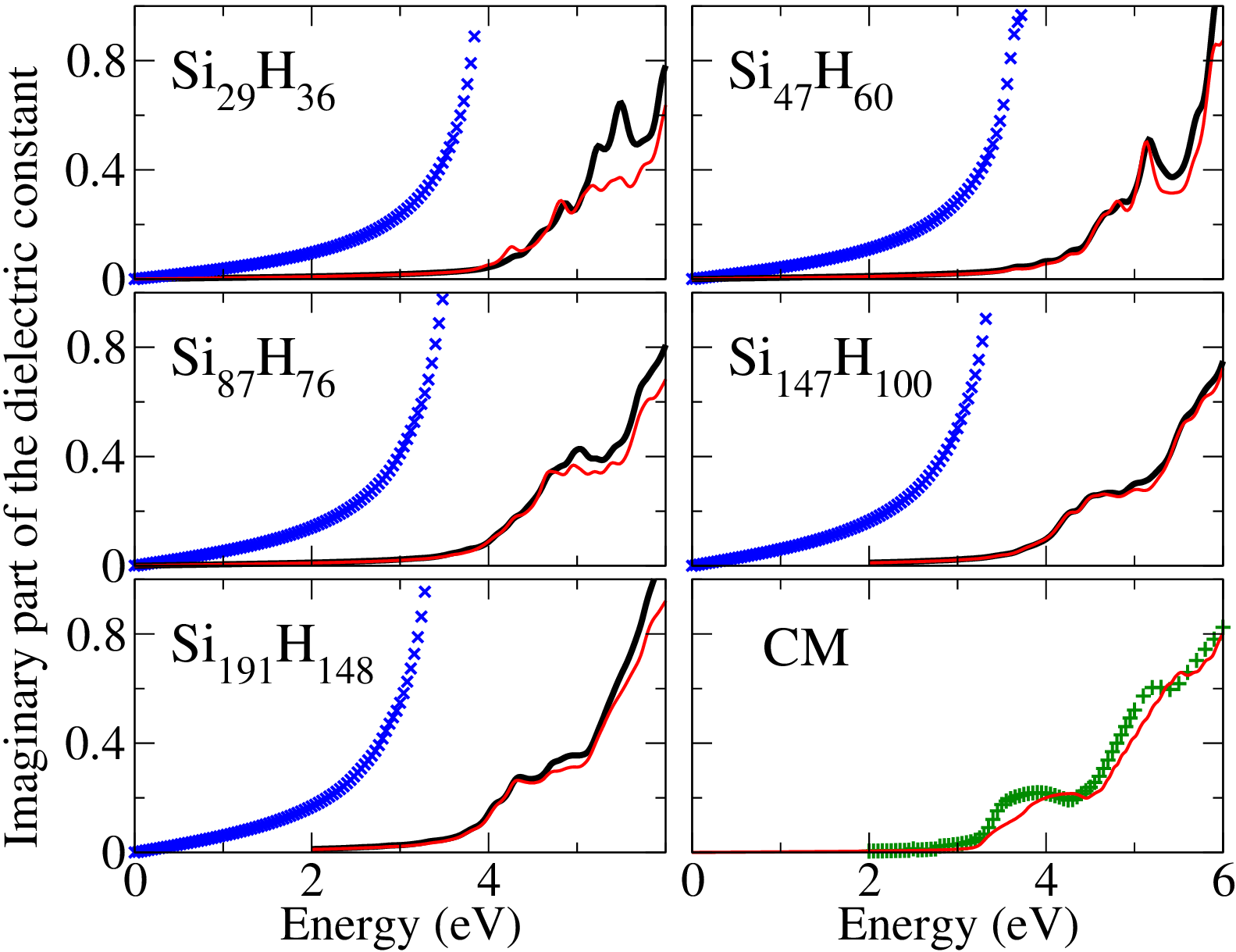}
	\caption{(Color online) Imaginary part of the dielectric constant for a set of increasing size Si-\textit{nc}. RPA, blue cross symbols; RPA+LF, black thick line; SCM, red thin line. For large nanocrystals, quantum effects become negligible. CM results are represented in the bottom part of the right-hand panel. They are obtained by using the bulk Si RPA dielectric constant (red-thin line), or experimental data (green-plus symbols).}
	\label{fig:fig2}
\end{figure}

Figure \ref{fig:fig2} shows the imaginary part of the frequency dependent effective dielectric constant for several Si-\textit{nc} and the classical model. A comparison between RPA, RPA+LF, and SCM is done on increasing the size. CM curves are calculated using the generalization to a dispersive medium of Eq. (\ref{eq:classical}), in which $\varepsilon_{\text{Si}}$ is either the RPA (red thin line) or the experimental data (green plus symbols). The agreement between the two curves is good.
As discussed above for the static dielectric constant, LFEs are mostly due to SPEs, and the agreement between RPA+LF and SCM improves on increasing the Si-\textit{nc} size. Furthermore, the convergence with the size is quite fast, and the Si$_{147}$H$_{100}$, Si$_{191}$H$_{148}$, and bulk Si spectra have the same trend in the low energy range, the only relevant difference being an overall redshift due to QCEs. From Fig. \ref{fig:fig2} it comes out that SPEs lead to a strong suppression of absorption in the visible range. It is worth stressing that the Si-\textit{nc} optical absorption dramatically change when SPEs are taken into account. While RPA curves have structures quite similar to the bulk Si absorption spectrum, with several peaks due to the most relevant interband transition energies,\cite{trani05} RPA+LF curves have a monotonically increasing  behavior in the whole energy range considered here.

\begin{figure}[t]
 	\includegraphics[width=0.4\textwidth]{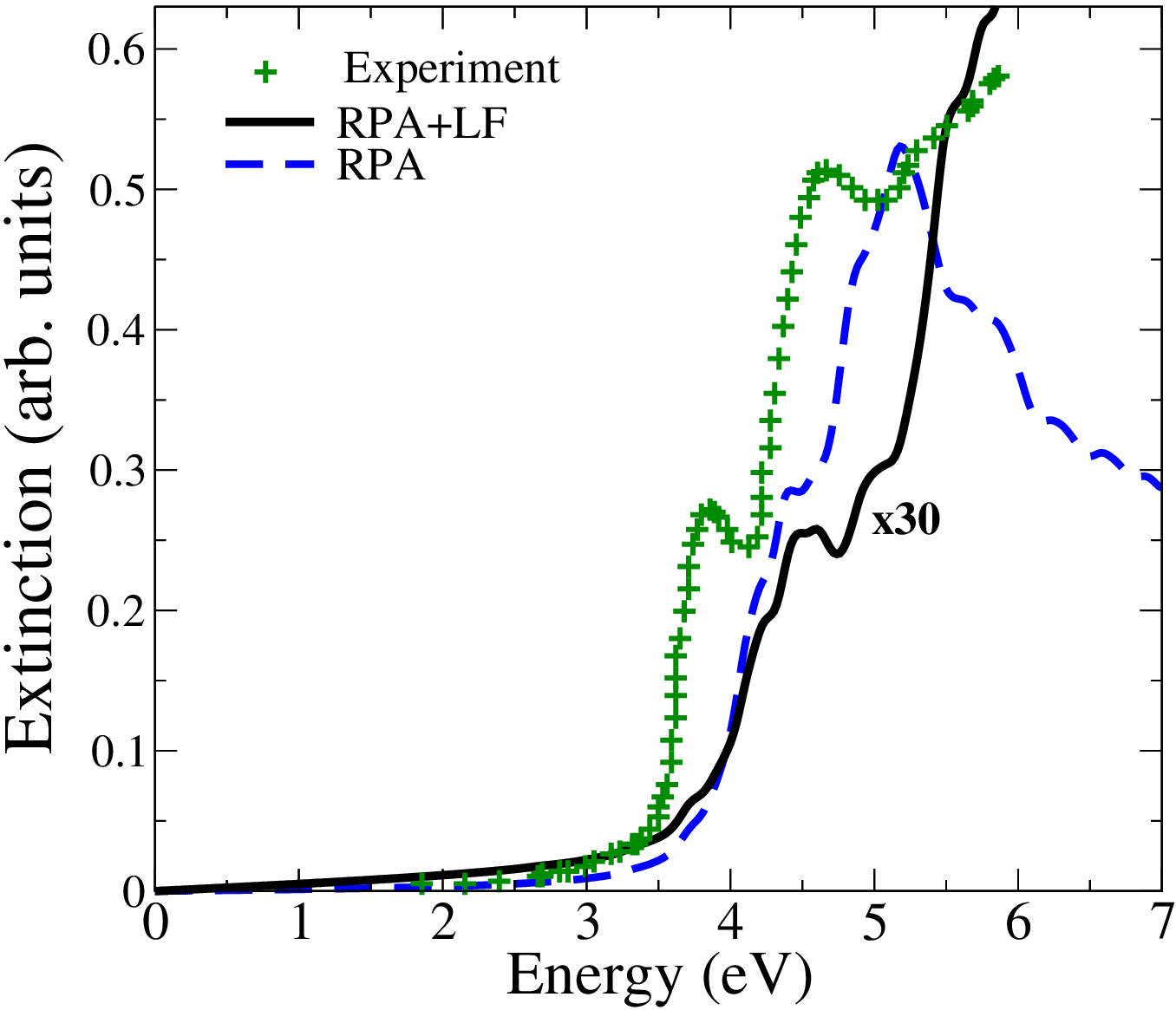}
	\caption{(Color online) Extinction coefficient of a d=1.8 nm Si-\textit{nc}. RPA (blue dashed line) and RPA+LF (black solid line) are compared with the experimental data, taken from Ref. \onlinecite{wilcoxon99}  (green symbols).}
	\label{fig:fig3}
\end{figure}

This behavior has been nicely confirmed by experiments. Figure \ref{fig:fig3} shows a comparison between the measured and the computed extinction coefficient of a Si nc with a size of 1.8 nm (green symbols). We have already compared the RPA calculation (blue dashed curve) with this experimental datum.\cite{trani05} Here we discuss the role of local fields, by a direct comparison of the previous results with the RPA+LF curve, shown by a black solid line in Fig. \ref{fig:fig3}. From the comparison, we can see that while the RPA curve decreases to zero after a main peak at about 5 eV, the experimental curve shows an increasing behavior with a good agreement with the RPA+LF curve.

It is important to remark that SPEs play a key role for the physical understanding of the spectra calculated by the time dependent density functional theory (TDDFT). When TDDFT absorption spectra were calculated for Si-\textit{nc}, a strong discrepancy with respect to the DFT result came out.\cite{vasiliev01} Recently, an interpretation of TDDFT results based on the classical theory has been given.\cite{tsolakidis05, idrobo06}
The DFT absorption spectra have peaks centered on the energies relative to the most relevant filled-to-empty state electronic transitions, and they are in nice agreement with the RPA calculations performed with semiempirical methods.\cite{notacomp}
The situation is very different when TDDFT calculations are performed. Si-\textit{nc} acquire a featureless spectrum in the low energy range (0-6 eV), showing a nearly monotonically  increasing behavior.\cite{vasiliev01}
Recently, for small Si-\textit{nc} it has been shown  that TDDFT gives results similar to RPA+LF.\cite{gatti05,sottile05}
\begin{figure}[t]
 	\includegraphics[width=0.4\textwidth]{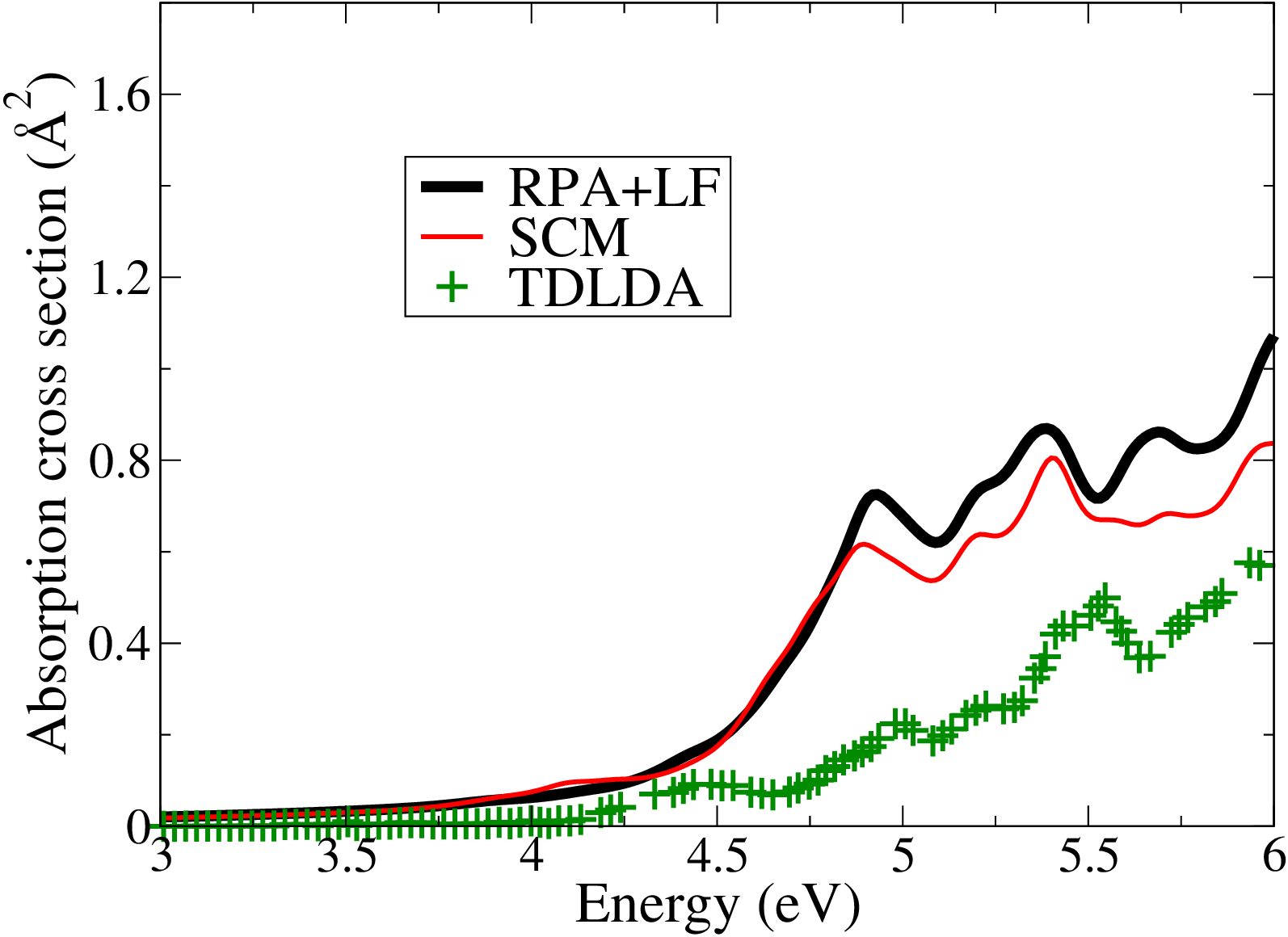}
	\caption{(Color online) Absorption cross section of Si$_{35}$H$_{36}$. RPA+LF results, black thick line; semiclassical, red thin line; TDLDA results, green symbols. TDLDA results are taken from Ref. \onlinecite{benedict03}.}
	\label{fig:fig4}
\end{figure}
In Fig \ref{fig:fig4} a comparison of the present RPA+LF results with the TDLDA calculations of Ref. \onlinecite{benedict03} is shown. By considering the huge difference between the two schemes, the agreement is fair.

In this paper, local field effects in the optical response of silicon nanocrystals have been studied. They have been shown to dramatically influence both the absorption spectra and the static dielectric constant. The cause lies in the surface polarization effects that is a key ingredient in the study of finite systems.\cite{resta06} We have shown that the dielectric model well reproduces the atomistic results. Moreover, we have shown that the comparison with experiments and first principles results is good.

The authors are grateful to Christophe Delerue, for very fruitful discussions and a critical reading of the paper. Financial support by COFIN-PRIN 2005 is acknowledged. Calculations have been performed at CINECA-``Progetti Supercalcolo 2006'' and ``Campus Computational Grid''-Universit\`a di Napoli ``Federico II'' advanced computing facilities.

% \bibliographystyle{apsrev}
% \bibliography{trani}

\newpage
\begin{thebibliography}{26}
\expandafter\ifx\csname natexlab\endcsname\relax\def\natexlab#1{#1}\fi
\expandafter\ifx\csname bibnamefont\endcsname\relax
  \def\bibnamefont#1{#1}\fi
\expandafter\ifx\csname bibfnamefont\endcsname\relax
  \def\bibfnamefont#1{#1}\fi
\expandafter\ifx\csname citenamefont\endcsname\relax
  \def\citenamefont#1{#1}\fi
\expandafter\ifx\csname url\endcsname\relax
  \def\url#1{\texttt{#1}}\fi
\expandafter\ifx\csname urlprefix\endcsname\relax\def\urlprefix{URL }\fi
\providecommand{\bibinfo}[2]{#2}
\providecommand{\eprint}[2][]{\url{#2}}

\bibitem[{\citenamefont{Wilson et~al.}(1993)\citenamefont{Wilson, Szajowski,
  and Brus}}]{wilson93}
\bibinfo{author}{\bibfnamefont{W.~L.} \bibnamefont{Wilson}},
  \bibinfo{author}{\bibfnamefont{P.~F.} \bibnamefont{Szajowski}},
  \bibnamefont{and} \bibinfo{author}{\bibfnamefont{L.~E.} \bibnamefont{Brus}},
  \bibinfo{journal}{Science} \textbf{\bibinfo{volume}{262}},
  \bibinfo{pages}{1242} (\bibinfo{year}{1993}).

\bibitem[{\citenamefont{Wilcoxon et~al.}(1999)\citenamefont{Wilcoxon, Samara,
  and Provencio}}]{wilcoxon99}
\bibinfo{author}{\bibfnamefont{J.~P.} \bibnamefont{Wilcoxon}},
  \bibinfo{author}{\bibfnamefont{G.~A.} \bibnamefont{Samara}},
  \bibnamefont{and} \bibinfo{author}{\bibfnamefont{P.~N.}
  \bibnamefont{Provencio}}, \bibinfo{journal}{Phys. Rev. B}
  \textbf{\bibinfo{volume}{60}}, \bibinfo{pages}{2704} (\bibinfo{year}{1999}).

\bibitem[{\citenamefont{Warner et~al.}(2005)\citenamefont{Warner, Hoshino,
  Yamamoto, and Tilley}}]{warner05}
\bibinfo{author}{\bibfnamefont{J.~H.} \bibnamefont{Warner}},
  \bibinfo{author}{\bibfnamefont{A.}~\bibnamefont{Hoshino}},
  \bibinfo{author}{\bibfnamefont{K.}~\bibnamefont{Yamamoto}}, \bibnamefont{and}
  \bibinfo{author}{\bibfnamefont{R.~D.} \bibnamefont{Tilley}},
  \bibinfo{journal}{Angew. Chem., Int. Ed.} \textbf{\bibinfo{volume}{44}},
  \bibinfo{pages}{4550} (\bibinfo{year}{2005}).

\bibitem[{\citenamefont{Jurbergs et~al.}(2006)\citenamefont{Jurbergs, Rogojina,
  Mangolini, and Kortshagen}}]{jurbergs06}
\bibinfo{author}{\bibfnamefont{D.}~\bibnamefont{Jurbergs}},
  \bibinfo{author}{\bibfnamefont{E.}~\bibnamefont{Rogojina}},
  \bibinfo{author}{\bibfnamefont{L.}~\bibnamefont{Mangolini}},
  \bibnamefont{and}
  \bibinfo{author}{\bibfnamefont{U.}~\bibnamefont{Kortshagen}},
  \bibinfo{journal}{Appl. Phys. Lett.} \textbf{\bibinfo{volume}{88}},
  \bibinfo{pages}{233116} (\bibinfo{year}{2006}).

\bibitem[{\citenamefont{Puzder et~al.}(2002)\citenamefont{Puzder, Williamson,
  Grossman, and Galli}}]{puzder02}
\bibinfo{author}{\bibfnamefont{A.}~\bibnamefont{Puzder}},
  \bibinfo{author}{\bibfnamefont{A.~J.} \bibnamefont{Williamson}},
  \bibinfo{author}{\bibfnamefont{J.~C.} \bibnamefont{Grossman}},
  \bibnamefont{and} \bibinfo{author}{\bibfnamefont{G.}~\bibnamefont{Galli}},
  \bibinfo{journal}{Phys. Rev. Lett.} \textbf{\bibinfo{volume}{88}},
  \bibinfo{pages}{097401} (\bibinfo{year}{2002}).

\bibitem[{sig()}]{sigle}
\bibinfo{note}{In order to simplify the reading, we adopt the following
  abbreviations: RPA (random phase approximation), LFE (local field effects),
  SPE (surface polarization effects), QCE (quantum confinement effect), CM
  (classical model), SCM (semiclassical model).}

\bibitem[{\citenamefont{Trani et~al.}(2005)\citenamefont{Trani, Cantele, Ninno,
  and Iadonisi}}]{trani05}
\bibinfo{author}{\bibfnamefont{F.}~\bibnamefont{Trani}},
  \bibinfo{author}{\bibfnamefont{G.}~\bibnamefont{Cantele}},
  \bibinfo{author}{\bibfnamefont{D.}~\bibnamefont{Ninno}}, \bibnamefont{and}
  \bibinfo{author}{\bibfnamefont{G.}~\bibnamefont{Iadonisi}},
  \bibinfo{journal}{Phys. {R}ev. {B}} \textbf{\bibinfo{volume}{72}},
  \bibinfo{pages}{075423} (\bibinfo{year}{2005}).

\bibitem[{\citenamefont{Landau and Lifshitz}(1984)}]{landau}
\bibinfo{author}{\bibfnamefont{L.~D.} \bibnamefont{Landau}} \bibnamefont{and}
  \bibinfo{author}{\bibfnamefont{E.~M.} \bibnamefont{Lifshitz}},
  \emph{\bibinfo{title}{Electrodynamics of Continuous Media}},
  \bibinfo{edition}{2nd} ed. (\bibinfo{publisher}{Pergamon}, \bibinfo{address}{Oxford},
  \bibinfo{year}{1984}).

\bibitem[{\citenamefont{Wang and Zunger}(1994)}]{wang94}
\bibinfo{author}{\bibfnamefont{L.~W.} \bibnamefont{Wang}} \bibnamefont{and}
  \bibinfo{author}{\bibfnamefont{A.}~\bibnamefont{Zunger}},
  \bibinfo{journal}{Phys. {R}ev. {L}ett.} \textbf{\bibinfo{volume}{73}},
  \bibinfo{pages}{1039} (\bibinfo{year}{1994}).

\bibitem[{\citenamefont{Lannoo et~al.}(1995)\citenamefont{Lannoo, Delerue, and
  Allan}}]{lannoo95}
\bibinfo{author}{\bibfnamefont{M.}~\bibnamefont{Lannoo}},
  \bibinfo{author}{\bibfnamefont{C.}~\bibnamefont{Delerue}}, \bibnamefont{and}
  \bibinfo{author}{\bibfnamefont{G.}~\bibnamefont{Allan}},
  \bibinfo{journal}{Phys. {R}ev. {L}ett.} \textbf{\bibinfo{volume}{74}},
  \bibinfo{pages}{3415} (\bibinfo{year}{1995}).

\bibitem[{not({\natexlab{a}})}]{nota1}
\bibinfo{note}{The polarizability of a dielectric sphere with dielectric
  constant $\varepsilon$ and volume $\Omega$, embedded into a background with
  dielectric constant $\varepsilon_0$, is given by
  $\alpha=3\Omega\varepsilon_0(\varepsilon-\varepsilon_0)/[4\pi(\varepsilon+2\varepsilon_0)]$.}

\bibitem[{\citenamefont{Delerue and Lannoo}(2004)}]{deleruebook}
\bibinfo{author}{\bibfnamefont{C.}~\bibnamefont{Delerue}} \bibnamefont{and}
  \bibinfo{author}{\bibfnamefont{M.}~\bibnamefont{Lannoo}},
  \emph{\bibinfo{title}{Nanostructures - Theory and Modelling}}
  (\bibinfo{publisher}{Springer-Verlag}, \bibinfo{address}{Berlin}, \bibinfo{year}{2004}).

\bibitem[{\citenamefont{Kovalev et~al.}(1995)\citenamefont{Kovalev, Ben-Chorin,
  Diener, Koch, Efros, Rosen, Gippius, and Tikhodeev}}]{kovalev95}
\bibinfo{author}{\bibfnamefont{D.}~\bibnamefont{Kovalev}},
  \bibinfo{author}{\bibfnamefont{M.}~\bibnamefont{Ben-Chorin}},
  \bibinfo{author}{\bibfnamefont{J.}~\bibnamefont{Diener}},
  \bibinfo{author}{\bibfnamefont{F.}~\bibnamefont{Koch}},
  \bibinfo{author}{\bibfnamefont{A.~L.} \bibnamefont{Efros}},
  \bibinfo{author}{\bibfnamefont{M.}~\bibnamefont{Rosen}},
  \bibinfo{author}{\bibfnamefont{N.~A.} \bibnamefont{Gippius}},
  \bibnamefont{and} \bibinfo{author}{\bibfnamefont{S.~G.}
  \bibnamefont{Tikhodeev}}, \bibinfo{journal}{Appl. Phys. Lett.}
  \textbf{\bibinfo{volume}{67}}, \bibinfo{pages}{1585} (\bibinfo{year}{1995}).

\bibitem[{\citenamefont{Gatti and Onida}(2005)}]{gatti05}
\bibinfo{author}{\bibfnamefont{M.}~\bibnamefont{Gatti}} \bibnamefont{and}
  \bibinfo{author}{\bibfnamefont{G.}~\bibnamefont{Onida}},
  \bibinfo{journal}{Phys. Rev. B} \textbf{\bibinfo{volume}{72}},
  \bibinfo{pages}{045442} (\bibinfo{year}{2005}).

\bibitem[{\citenamefont{Leung and Whaley}(1997)}]{leung97}
\bibinfo{author}{\bibfnamefont{K.}~\bibnamefont{Leung}} \bibnamefont{and}
  \bibinfo{author}{\bibfnamefont{K.~B.} \bibnamefont{Whaley}},
  \bibinfo{journal}{Phys. Rev. B} \textbf{\bibinfo{volume}{56}},
  \bibinfo{pages}{7455} (\bibinfo{year}{1997}).

\bibitem[{\citenamefont{Trani et~al.}(2006)\citenamefont{Trani, Ninno, Cantele,
  Hameeuw, Iadonisi, Degoli, and Ossicini}}]{trani06}
\bibinfo{author}{\bibfnamefont{F.}~\bibnamefont{Trani}},
  \bibinfo{author}{\bibfnamefont{D.}~\bibnamefont{Ninno}},
  \bibinfo{author}{\bibfnamefont{G.}~\bibnamefont{Cantele}},
  \bibinfo{author}{\bibfnamefont{K.~J.} \bibnamefont{Hameeuw}},
  \bibinfo{author}{\bibfnamefont{G.}~\bibnamefont{Iadonisi}},
  \bibinfo{author}{\bibfnamefont{E.}~\bibnamefont{Degoli}}, \bibnamefont{and}
  \bibinfo{author}{\bibfnamefont{S.}~\bibnamefont{Ossicini}},
  \bibinfo{journal}{Phys. Rev. B} \textbf{\bibinfo{volume}{73}},
  \bibinfo{pages}{245430} (\bibinfo{year}{2006}).

\bibitem[{\citenamefont{Bruneval et~al.}(2005)\citenamefont{Bruneval, Botti,
  and Reining}}]{bruneval05}
\bibinfo{author}{\bibfnamefont{F.}~\bibnamefont{Bruneval}},
  \bibinfo{author}{\bibfnamefont{S.}~\bibnamefont{Botti}}, \bibnamefont{and}
  \bibinfo{author}{\bibfnamefont{L.}~\bibnamefont{Reining}},
  \bibinfo{journal}{Phys. Rev. Lett.} \textbf{\bibinfo{volume}{94}},
  \bibinfo{pages}{219701} (\bibinfo{year}{2005}).

\bibitem[{\citenamefont{Sottile et~al.}(2005)\citenamefont{Sottile, Bruneval,
  Marinopoulos, Dash, Botti, Olevano, Vast, Rubio, and Reining}}]{sottile05}
\bibinfo{author}{\bibfnamefont{F.}~\bibnamefont{Sottile}},
  \bibinfo{author}{\bibfnamefont{F.}~\bibnamefont{Bruneval}},
  \bibinfo{author}{\bibfnamefont{A.~G.} \bibnamefont{Marinopoulos}},
  \bibinfo{author}{\bibfnamefont{L.~K.} \bibnamefont{Dash}},
  \bibinfo{author}{\bibfnamefont{S.}~\bibnamefont{Botti}},
  \bibinfo{author}{\bibfnamefont{V.}~\bibnamefont{Olevano}},
  \bibinfo{author}{\bibfnamefont{N.}~\bibnamefont{Vast}},
  \bibinfo{author}{\bibfnamefont{A.}~\bibnamefont{Rubio}}, \bibnamefont{and}
  \bibinfo{author}{\bibfnamefont{L.}~\bibnamefont{Reining}},
  \bibinfo{journal}{Int. J. Quantum Chem.} \textbf{\bibinfo{volume}{102}},
  \bibinfo{pages}{684} (\bibinfo{year}{2005}).

\bibitem[{\citenamefont{Delerue et~al.}(1997)\citenamefont{Delerue, Lannoo, and
  Allan}}]{delerue97}
\bibinfo{author}{\bibfnamefont{C.}~\bibnamefont{Delerue}},
  \bibinfo{author}{\bibfnamefont{M.}~\bibnamefont{Lannoo}}, \bibnamefont{and}
  \bibinfo{author}{\bibfnamefont{G.}~\bibnamefont{Allan}},
  \bibinfo{journal}{Phys. Rev. B} \textbf{\bibinfo{volume}{56}},
  \bibinfo{pages}{15306} (\bibinfo{year}{1997}).

\bibitem[{\citenamefont{Hanke and Sham}(1974)}]{hanke74}
\bibinfo{author}{\bibfnamefont{W.}~\bibnamefont{Hanke}} \bibnamefont{and}
  \bibinfo{author}{\bibfnamefont{L.}~\bibnamefont{Sham}},
  \bibinfo{journal}{Phys. Rev. Lett.} \textbf{\bibinfo{volume}{33}},
  \bibinfo{pages}{582} (\bibinfo{year}{1974}).

\bibitem[{\citenamefont{Benedict et~al.}(2003)\citenamefont{Benedict, Puzder,
  Williamson, Grossman, Galli, Klepeis, Raty, and Pankratov}}]{benedict03}
\bibinfo{author}{\bibfnamefont{L.~X.} \bibnamefont{Benedict}},
  \bibinfo{author}{\bibfnamefont{A.}~\bibnamefont{Puzder}},
  \bibinfo{author}{\bibfnamefont{A.~J.} \bibnamefont{Williamson}},
  \bibinfo{author}{\bibfnamefont{J.~C.} \bibnamefont{Grossman}},
  \bibinfo{author}{\bibfnamefont{G.}~\bibnamefont{Galli}},
  \bibinfo{author}{\bibfnamefont{J.~E.} \bibnamefont{Klepeis}},
  \bibinfo{author}{\bibfnamefont{J.~Y.} \bibnamefont{Raty}}, \bibnamefont{and}
  \bibinfo{author}{\bibfnamefont{O.}~\bibnamefont{Pankratov}},
  \bibinfo{journal}{Phys. Rev. B} \textbf{\bibinfo{volume}{68}},
  \bibinfo{pages}{085310} (\bibinfo{year}{2003}).

\bibitem[{\citenamefont{Vasiliev et~al.}(2001)\citenamefont{Vasiliev, {\"O}{\u
  g}{\"u}t, and Chelikowsky}}]{vasiliev01}
\bibinfo{author}{\bibfnamefont{I.}~\bibnamefont{Vasiliev}},
  \bibinfo{author}{\bibfnamefont{S.}~\bibnamefont{{\"O}{\u g}{\"u}t}},
  \bibnamefont{and} \bibinfo{author}{\bibfnamefont{J.~R.}
  \bibnamefont{Chelikowsky}}, \bibinfo{journal}{Phys. Rev. Lett.}
  \textbf{\bibinfo{volume}{86}}, \bibinfo{pages}{1813} (\bibinfo{year}{2001}).

\bibitem[{\citenamefont{Tsolakidis and Martin}(2005)}]{tsolakidis05}
\bibinfo{author}{\bibfnamefont{A.}~\bibnamefont{Tsolakidis}} \bibnamefont{and}
  \bibinfo{author}{\bibfnamefont{R.~M.} \bibnamefont{Martin}},
  \bibinfo{journal}{Phys. Rev. B} \textbf{\bibinfo{volume}{71}},
  \bibinfo{pages}{125319} (\bibinfo{year}{2005}).

\bibitem[{\citenamefont{Idrobo et~al.}(2006)\citenamefont{Idrobo, Yang,
  Jackson, and {\"O}{\u{g}}{\"u}t}}]{idrobo06}
\bibinfo{author}{\bibfnamefont{J.~C.} \bibnamefont{Idrobo}},
  \bibinfo{author}{\bibfnamefont{M.}~\bibnamefont{Yang}},
  \bibinfo{author}{\bibfnamefont{K.~A.} \bibnamefont{Jackson}},
  \bibnamefont{and}
  \bibinfo{author}{\bibfnamefont{S.}~\bibnamefont{{\"O}{\u{g}}{\"u}t}},
  \bibinfo{journal}{Phys. Rev. B} \textbf{\bibinfo{volume}{74}},
  \bibinfo{pages}{153410} (\bibinfo{year}{2006}).

\bibitem[{not({\natexlab{b}})}]{notacomp}
\bibinfo{note}{A remarkable proof of the close similarity between DFT and
  semiempirical RPA curves comes from the comparison of the spectra shown in
  Refs. \onlinecite{trani05} and \onlinecite{vasiliev01}, especially for
  Si$_{147}$H$_{148}$}.

\bibitem[{\citenamefont{Resta}(2006)}]{resta06}
\bibinfo{author}{\bibfnamefont{R.}~\bibnamefont{Resta}},
  \bibinfo{journal}{Phys. Rev. Lett.} \textbf{\bibinfo{volume}{96}},
  \bibinfo{pages}{137601} (\bibinfo{year}{2006}).

\end{thebibliography}

\end{document}